\documentclass[twocolumn,superscriptaddress,floatfix,amsmath,footinbib,amssymb]{revtex4-1}
\usepackage{amssymb}
\usepackage{amsmath}
\usepackage{epsfig}
\usepackage{t1enc}
\usepackage{soul}
\usepackage{color}

\begin{document}
\newcommand{\beq}{\begin{equation}}
\newcommand{\eeq}{\end{equation}}
\newcommand{\bea}{\begin{eqnarray}}
\newcommand{\eea}{\end{eqnarray}}
\title{Localised projective measurement of a quantum field in non-inertial frames}

\author{Andrzej Dragan}
\affiliation{School of Mathematical Sciences, University of Nottingham, Nottingham NG7 2RD, United Kingdom}
\address{Institute of Theoretical Physics, University of Warsaw, Ho\.{z}a 69, 00-049 Warsaw, Poland}
\author{Jason Doukas}
\address{National Institute for Informatics, 2-1-2 Hitotsubashi, Chiyoda-ku, Tokyo 101-8430,  Japan}
\author{Eduardo Mart\'{i}n-Mart\'{i}nez}
\affiliation{Institute for Quantum Computing, Department of Physics and Astronomy and Department of Applied Mathematics, University of Waterloo, 200 University
Avenue W, Waterloo, Ontario, N2L 3G1, Canada}
\author{David Edward Bruschi}
\affiliation{School of Mathematical Sciences, University of Nottingham, Nottingham NG 72RD, United Kingdom}
\date{\today}

\begin{abstract}
We propose a projective operator formalism that is well-suited to study the correlations of quantum fields in non-inertial frames.  We generalise a Glauber model of detection of a single localised field mode that is capable of making measurements in an arbitrary reference frame. We show that the model correctly reproduces the Unruh temperature formula of a single accelerated detector, and use it to extract vacuum entanglement by a pair of counter-accelerating detectors. This latter example is a proof of principle that this approach will be appropriate to further studies on the nature of entanglement in non-inertial frames and, in general, to model experimentally feasible scenarios in quantum field theory in non-inertial frames. Finally, as further confirmation of the validity of our approach, we introduce an explicit perturbative matter-radiation interaction model which reproduces both the generalised Glauber model and the projective measurement results in the weak coupling regime.
\end{abstract}

\maketitle

\section{Introduction}
One of the most fascinating features of relativistic quantum fields is that the notion of a particle is frame dependent. Even if a given observer describes the field state as free from particles, the state of the same field described by another observer can be in fact populated by particles. This relativistic effect is a direct consequence of the Bogolyubov transformation between inertial and non-inertial reference frames and has inspired the development of the whole new field of relativistic quantum information. The main objective of this study is to find the consequences of relativistic effects on protocols describing the storage, processing and transmission of quantum information. Perhaps most striking of these investigations has been the realisation that entanglement, which is a key resource in quantum information protocols, depends on the state of motion of the observer. 

Seminal works on this topic \cite{Alsing2003} studied the effects of acceleration on the entanglement of global states of free fields. However, several issues remained unsettled, including how to properly describe a possible experimental setting \cite{Schutzhold2005} and how to correctly address the technical solutions beyond the single mode approximation \cite{Bruschi2010}. The recent approach \cite{Bruschi2010} reanalysed the general setting of \cite{Alsing2003} beyond the single mode approximation. However, it required the preparation of a whole family of orthogonal states of global Unruh modes, one for each accelerated observer, instead of a single, fixed state given to all observers. It is was also not clear how such global modes could be physically prepared. For these reasons the physical effects of acceleration on entanglement remained in disrepute.  

In this paper we describe a framework for which these issues can be resolved. We introduce a model of a detector that performs a projective measurement on a localised single-mode of the quantum field. Our detector can either move inertially or accelerate. This framework allows a fixed state of the field defined in an inertial frame to be probed for any acceleration. Thereby avoiding having to use a different initial state for each acceleration and a non-local initial mode.  

To motivate the projective measurement formalism we introduce a relativistic generalisation of the Glauber model \cite{Glauber1963} and show the relationship between the quantities calculated in each approach. One particularly nice feature of the Glauber detector is that the detector only clicks when particles are present in the mode that the device measures. This lends itself to a \textit{particle} detector interpretation rather than a detector of fluctuations like that of the Unruh-DeWitt detector \cite{Letaw1980}. 

The model is applied to the simplest case of the Minkowski vacuum state and we use it to reproduce the well-known Unruh effect for a single accelerated detector \cite{Unruh1976}.  We then study how a pair of such detectors can extract nonlocal correlations present in the vacuum \cite{Reznik2003}, showing that our model is suitable for studies of entanglement in non-inertial frames. Our analysis can be extended to arbitrary states of the field, which is the subject of another work \cite{Dragan2012}.

We can freely choose the size and the shape of the detected wave-packet mode allowing one to study how these properties affect the measurement results. We find that the entanglement extracted from the vacuum by a pair of uniformly counter-accelerating detectors increases with the size of the wave-packet mode and the detector's acceleration.

We first outline, in section \ref{sec:MoD}, the conceptual ideas behind the detector model presented here. Then, in section \ref{sec:Not} we introduce the formalism to work with uniformly accelerated observers of a quantum field and discuss the process of field quantisation in the Rindler basis. We  give then a heuristic argument for a natural choice of physically meaningful localised detector modes in section \ref{sec:QoSF}.   In section \ref{sec:particles} we calculate the average number of particles populating the field mode probed by a localised accelerated detector when the field state is the Minkowski vacuum state; and in section  \ref{sec:Projective} we introduce the formalism for the projective measurements on the field mode probed by the detector. This will allow us to estimate the temperature of the photo-count statistics (in section \ref{sec:Unruh}) and  to study the extraction of field entanglement of spatially separated spacetime regions (in section \ref{sec:vacent}).  Our conclusions are presented in section \ref{sec:conclusion}. Additionally, a detailed analysis on the photo-detector interaction dynamics is given in appendix \ref{appendix1}.

\section{Model of the detector}\label{sec:MoD}
We consider a model of measurement carried out locally on a quantum field with a device that is only sensitive to a single localised wave-packet mode. 
\begin{figure}
\begin{center}
\includegraphics[width=.35\textwidth]{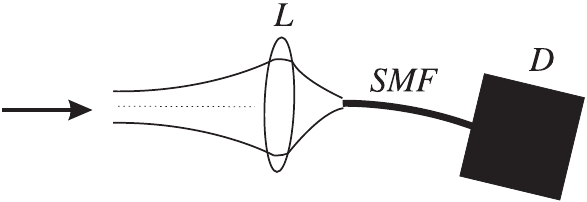}
\end{center}
\caption{\label{scheme}Detection of a quantum state of a single light-mode by means of a lens $L$, single mode fibre $SMF$ and a photon-counting detector $D$.}
\end{figure}
The model is inspired by a quantum-optical setting depicted in Fig.~\ref{scheme} that has been considered both in theory \cite{Dragan2004} and realised experimentally \cite{Banaszek2004}. The original scheme consists of a lens, $L$, focusing incident light onto one end of a single-mode optical fibre $SMF$ that transmits only one particular transverse mode of light. The other end of the fibre is attached to a single-photon detector, $D$, utilising the photoelectric effect to detect and count single light quanta arriving in the fibre-coupled mode of light. In the idealised version of this scheme the state of the single field-mode is projected onto the Fock basis and all the other modes that are orthogonal to the one transmitted through the fibre are not affected by the detection process. This scheme can also be generalised to allow the detection of an arbitrary mode by placing optical elements in front of the lens, $L$, which transform the given input mode into the mode transmitted by the fibre.

Notice that the state of the measured system is completely destroyed during the measurement process (photons are absorbed) and the physical measurement corresponding to this model is not instantaneous. The minimum time of such a detection is equal to the time it takes light collected by the lens to reach the SMF plus the time it takes for the device that captures the light to make the read-out. Therefore the bigger the mode, the longer it takes to extract it from all the other modes. In many practical scenarios, such as the one considered above, the time of detection can be safely neglected. However, one has to bear in mind that such idealisation has its limits and not taking measurement time into account can lead to problems \cite{Sorkin1993}. 

In this work we consider an abstract "black-box" device that measures a wave-packet mode of the field in the same way: the process consists of filtering out the relevant mode and then measuring its state. For simplicity we consider a 1+1 dimensional version of the above setting. The process we describe disturbs the field only in the proximity of the measurement device; we assume that the detector's response is not affected by the state of other modes orthogonal to the one it measures, and that the measurements performed on this mode do not affect the state of the orthogonal modes. In this sense the process is localised in space.

The quantum field under consideration is taken to be a real non-interacting massless scalar field, $\hat{\phi}$, described by a Klein-Gordon equation, $\square\hat{\phi} = 0$, in 1+1 Minkowski spacetime. Let the wave-packet observed by the detector be denoted by $\psi_\text{D}(\xi,\tau)$, where $\xi$ and $\tau$, are coordinates of the reference frame co-moving with the detector. We assume that $\psi_\text{D}(\xi,\tau)$ is consists only of positive frequency waves with respect to $\tau$. Although the splitting into positive and negative frequencies is frame dependent, the linear combination $\psi_\text{D}(\xi,\tau)+\psi_\text{D}^\star(\xi,\tau)$ does not depend on the coordinate system. It is this real field solution that describes the classical wave which couples to the detector. Using the positive frequency part it is possible to define an operator $\hat{d}=(\psi_\text{D},\hat{\phi})$, associated with the wave-packet that the detector sees. Here $(\cdot,\cdot)$ is the Klein-Gordon scalar product defined on the set of complex solutions to the Klein-Gordon equation by:
\bea
(\phi_1,\phi_2)=i\int dx \sqrt{-g}\phi_1^{\star}(x,t)\overleftrightarrow{\partial_t}\phi_2(x,t)|_{t=0},
\eea
where $g$ is the determinant of the metric and $\overleftrightarrow{\partial_t}$ is the two-edged differential operator defined by:  $f\overleftrightarrow{\partial_t} h=f\frac{\partial h}{c\partial t}-\frac{\partial f}{c\partial t}h$. If $\psi_\text{D}$ is unit normalised with respect to the Klein-Gordon scalar product, then $\hat{d}$ satisfies the usual commutation relations $[\hat{d},\hat{d}^{\dagger}]=1$. If the motion of the detector is inertial (uniformly accelerated) then $\hat{d}$ annihilates the Minkowski (Rindler) vacuum state. This leads to the identification of $\hat{d}^{(\dagger)}$ as an annihilation (creation) operator.  Any state of the field that is formed by the linear superposition of the basis states $\{\frac{\hat{d}^{\dagger n}}{\sqrt{n!}}|0\rangle\}$ or a mixture of them is said to be a state in the mode $\psi_D$. The operator $\hat{d}^{\dagger}\hat{d}$ is a number operator for the number of particles in the mode $\psi_\text{D}$. These excitations are, however, not energy eigenstates of the free-field hamiltonian. The state accessible to the detector is the total field state $|\Psi\rangle$ projected onto the $\psi_\text{D}$ subspace:
\bea
\label{mixedstate}
\text{Tr}_{\perp \psi_\text{D}} |\Psi\rangle \langle \Psi |,
\eea
where the trace is to be taken over the subspace of all modes orthogonal to $\psi_\text{D}$. In this work we assume that $\psi_\text{D}$ is localised  \cite{Referee1}, which will allow us to attribute the detector a certain position in space. For example if the measurement on $\psi_\text{D}$ takes place at times close to $\tau=0$ then the detector has to be placed in the region where $\psi_\text{D}(\xi,0)$ is non-zero.

One of the greatest advancements of the field of quantum optics occurred with the introduction of the Glauber model \cite{Glauber1963}. The detection process described in the Minkowski frame by the Glauber model for the scalar field results in a particle absorption occurring at the point $(ct, x)$, described in the weak coupling regime by the following absorption amplitude:
\bea\label{GlauberTransition}
A_{if}(x,t)=\langle f| \hat{\phi}^{+}(x,t)|i\rangle,
\eea
where $|i\rangle$ and $|f\rangle$ are initial and final states of the field and $\hat{\phi}^+(x,t)$ is the positive frequency part of the field operator in Minkowski coordinates. However, the final state of the field remains unknown after the measurement, so the total probability ${\cal P}$ of observing a particle is given by taking the squared modulus of the amplitude and summing over all final states. Generalising to mixed states and substituting \eqref{mixedstate} for the initial state results in:
\bea\label{GlauberMixedProb}
{\cal P}(x,t)=\text{Tr} \left\{\hat{\phi}^{-}(x,t)\hat{\phi}^{+}(x,t) \text{Tr}_{\perp \psi_\text{D}} |\Psi\rangle \langle \Psi |\right\}.
\eea

In this work we generalise the Glauber theory of detection to the case of detectors moving with relativistic accelerations. The crucial step in this generalisation is to replace the positive and negative frequency parts in the above equations with the splitting with respect to the proper time of the frame co-moving with the detector $(c\tau,\xi)$. Since the positive and negative frequency parts of the field are different in the accelerating and inertial frames, the intensity of particle clicks in the photo-counter will depend on the detector's motion. We prove in appendix \ref{appendix1} that our generalised Glauber model agrees to first order in perturbation theory with the predictions of a dynamical field-apparatus model that utilises an Unruh-DeWitt interaction, analogously to the inertial case \cite{Mandel}.  We find that even in the non-inertial case the measurement outcomes can be completely described by a set of projective operators, as in the standard quantum-optical non-relativistic setups.

\section{The accelerated frame of reference}\label{sec:Not}
\begin{figure}
\begin{center}
\includegraphics[width=.5\textwidth]{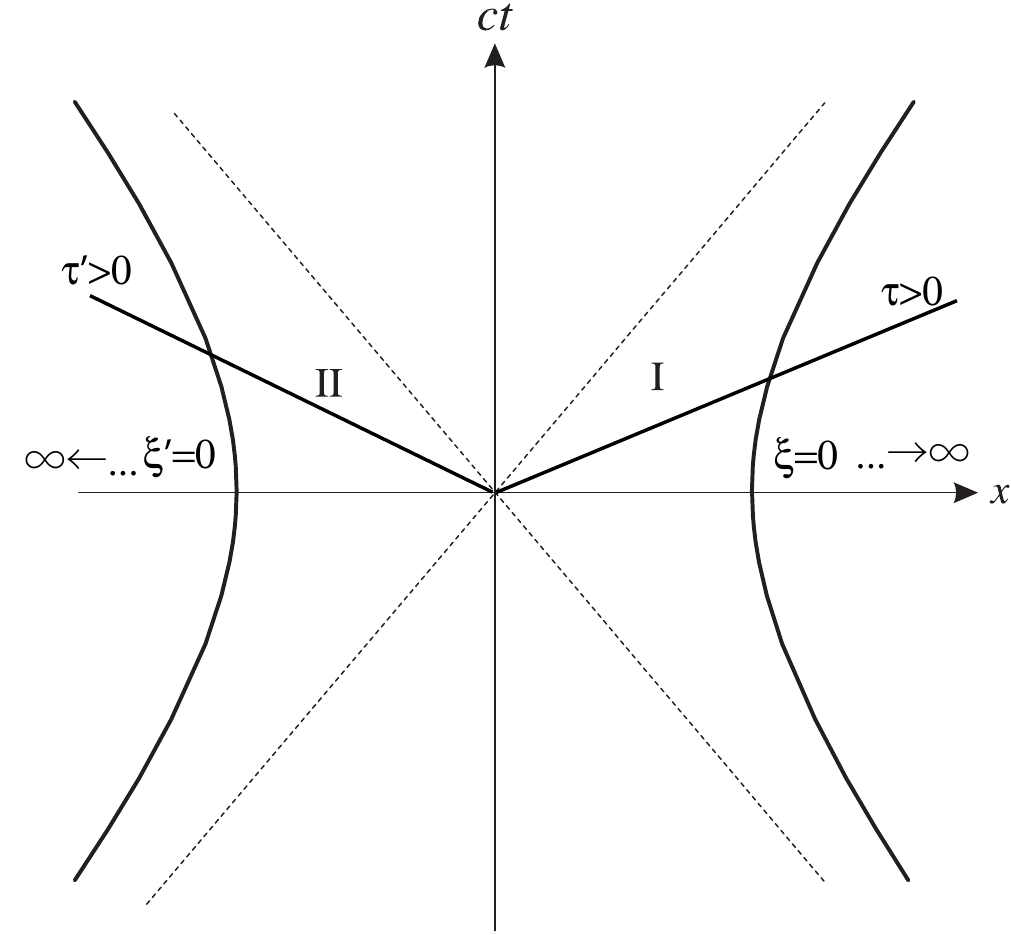}
\end{center}
\caption{\label{fig:Rindler} Minkowski diagram, with Rindler regions I $(x>c|t|)$ and II $(x<-c|t|)$ covered with conformal Rindler coordinates $(\tau,\xi)$ and $(\tau',\xi')$ respectively.}
\end{figure}

In this paper we focus exclusively on measurements made by the uniformly and relativistically accelerated detectors. A natural choice of coordinates for the accelerated observers are the Rindler coordinates:
\begin{eqnarray}
ct &=& \frac{c^2}{a}e^{a\xi/c^2} \sinh \frac{a\tau}{c}, ~~x = \frac{c^2}{a}e^{a\xi/c^2} \cosh \frac{a\tau}{c}, ~~(I)\nonumber \\ \nonumber \\
ct &=& \frac{c^2}{a}e^{a\xi'/c^2} \sinh \frac{a\tau'}{c}, ~~x = -\frac{c^2}{a}e^{a\xi'/c^2} \cosh \frac{a\tau'}{c}, ~~(II),\nonumber \\
\end{eqnarray}
where $-\infty<\xi,\xi', \tau, \tau'<\infty$, $a$ is an arbitrary parameter, and $I$, $II$ represent the two distinct regions of spacetime shown in Fig.~\ref{fig:Rindler}. To interpret $a$ as the proper acceleration of the detector, we must ensure that the photo-counter is spatially localised around $\xi = 0$ when $a>0$, or $\xi'=0$ when $a<0$. Then all of its components will approximatively experience the same proper time, which coincides with Rindler time $\tau$. 

One can decompose the field operator into a countable sum of Rindler modes by putting the field in each region into a box of length $h$ satisfying periodic boundary conditions. Then:
\begin{equation}
\label{rindlerdecomp}
\hat{\phi}=\sum_{k}\hat{b}_{k,I}w_{k,I}(\xi,\tau)+\hat{b}_{k,II}w_{k,II}(\xi',\tau')+h.c.,
\end{equation}
where $k=\frac{2 \pi n}{h}$ with $n=\pm1,\pm2\dots$, and
\bea\label{wdefI}
\sqrt{\frac{h}{2\pi}}w_{k,I}(\xi,\tau)&\equiv& \frac{1}{\sqrt{4 \pi |k|}}e^{i(k\xi-|k|c\tau)}\theta(x-c|t|),\\
\sqrt{\frac{h}{2\pi}}w_{k,II}(\xi',\tau')&\equiv& \frac{1}{\sqrt{4 \pi |k|}}e^{i(-k\xi'-|k|c\tau')}\theta(-x-c|t|).\nonumber
\eea
The Rindler modes are box normalised so that $(w_{k,I},w_{k',I})=\delta_{kk'}$. We will also find it useful to re-express our results in the continuum limit $h\rightarrow\infty$. This is done in the usual way by replacing $\frac{2\pi}{h}\sum_k\rightarrow\int dk$ and $\sqrt{\frac{h}{2\pi}}w_{k,I}\rightarrow w_{k,I}$ where in the continuum limit the Rindler modes are plane-wave normalised $(w_{k,I},w_{k',I})=\delta(k-k')$. The region I and II Rindler modes are positive frequency with respect to the Killing vectors $i\partial_{\tau}$ and $i\partial_{\tau'}$ in region I and II respectively, and have been chosen such that $k>0$ corresponds to a right moving wave; the asymmetry in sign between equations (\ref{wdefI}) occurs because $\xi$ increases in the right direction while $\xi'$ increases toward the left.

It is a known fact \cite{Unruh1976} that the Minkowski vacuum state can be expressed as a multi-frequency two-mode-squeezing of the Rindler vacuum state: 
\begin{equation}
\label{squeezedvacuum}
|0\rangle_{\text{M}} = \hat{S}_{I,II}|0\rangle_{\text{R}},
\end{equation}
where the squeezing operator, $\hat{S}_{I,II}$,  is characterised by the squeezing parameter, $r_k={\rm arctanh}(e^{-\pi |k| c^2/a})$, and fulfils the following relations:
\begin{equation}
\label{commut}
\hat{S}_{I,II}^\dagger\hat{b}_{l,I}\hat{S}_{I,II}=\cosh{r_l}\hat{b}_{l,I}+\sinh{r_l}\hat{b}^\dagger_{l,II}.
\end{equation}
We will find these commutation relations extremely useful when performing the algebraic manipulations of the following sections.

\section{Choice of the detector mode}\label{sec:QoSF}

The wave-packet corresponding to the localised field mode probed by the detector $\psi_\text{D}(\xi, \tau)$ must be spatially localised around $\xi = 0$ at the measurement time around $\tau=0$. In principle, any positive frequency localised wave-packet satisfying this condition will be sufficient, but our choice will be dictated by the convenience of further calculations. 

Consider a resting detector that selects $\psi_\text{D}(x,t)$ such that it has a central frequency coinciding with the $N$-th eigenmode of a cavity. We ask: what wave-packet $\psi_\text{D}(\xi,\tau)$ would a uniformly accelerated detector select if it was to operate on this same principle? In order to answer this question we compare the eigenmodes of an ideal cavity at rest and the eigenmodes of the same cavity in uniform acceleration \cite{Downes2011}. One infers that the central frequency of the mode $\psi_\text{D}(\xi,\tau)$ should be $\frac{Nc}{\sigma}$, where, $\sigma = \frac{2c^2}{a}\text{asinh}\left(\frac{aL}{2c^2}\right)$, $L$ is the proper length of the cavity and $N$ is a natural number corresponding to the chosen cavity eigenmode. We choose a Gaussian envelope of that mode, characterised by the width $\sigma$, to make it localised, however all our main results will not depend on the specific choice of that envelope:
\begin{eqnarray}
\label{modefunction}
{\psi}_{\text{D}}(\xi,0) &=& \frac{1}{\sqrt{N\sqrt{2\pi}}}\exp\left[-\frac{\xi^2}{\sigma^2} + i \frac{N}{\sigma}\xi\right],\nonumber \\ \nonumber \\
\partial_\tau{\psi}_{\text{D}}(\xi,0) &=& -i\frac{Nc}{\sigma}{\psi}_{\text{D}}(\xi,0),
\end{eqnarray}
The above expressions always have some contribution from negative frequency components, however these are negligible for $N\gg 1$ (actually $N>3$ is already enough). To get rid of them completely we introduce an infra-red cut-off $\Lambda$ to the considered mode:
\bea
\psi_\text{D}(\xi, \tau) \rightarrow \frac{\sum_{k>\Lambda}(w_{I,k},\psi_{\text{D}})w_{I,k}(\xi, \tau)}{\sqrt{\sum_{k>\Lambda}|(w_{I,k},\psi_{\text{D}})|^2}}.
\eea 
Note this then also ensures that the single particle states exist for this degree of localisation \cite{Referee1}. The device is defined to be only sensitive to the right-moving modes in region I  but this choice is arbitrary and can be changed if necessary. In the limit of small accelerations, the mode \eqref{modefunction} written in Minkowski coordinates reduces to $\psi_\text{D}(x,0) = \frac{1}{\sqrt{N\sqrt{2\pi}}}\exp\left[-\frac{x^2}{L^2} + i \frac{N}{L}x\right]$ up to a translation along $x$. It reproduces a regularised $N$-th mode of a resting cavity of a length $L$.

\section{Detection of particles from the Vacuum state}\label{sec:particles}
Consider measuring the occupation number operator of the transmitted localised mode, $\hat{d}^\dagger\hat{d}$. If at $\tau=0$ the measured field is in a Fock state $|n\rangle$ of a mode $\phi(\xi,\tau)$ that coincides at $\tau=0$ with $\psi_\text{D}$, i.e., $\phi(\xi,0)=\psi_\text{D}(\xi,0),~~\partial_\tau\phi(\xi,0)=\partial_\tau\psi_\text{D}(\xi,0)$ then the outcome of the detection will be a classical variable $n$.  

To generalise the result to an arbitrary state of the field it is useful to introduce the following decomposition of the field operator:
\begin{equation}
\label{ddecomp}
\hat{\phi}(\xi,0) = \hat{d}\psi_\text{D}(\xi,0)+ \hat{d}^\dagger\psi_\text{D}^\star(\xi,0)+\hat{\phi'},
\end{equation}
where $\hat{\phi'}$ is the remaining part of the mode decomposition containing all the modes orthogonal to $\psi_\text{D}$. In principle, a complete and orthonormal basis could be found for the Hilbert space by using Gram-Schmidt orthonormalisation on the positive frequency subspace of complex solutions beginning with the vector $\psi_\text{D}$. However, it will not be required to calculate such a basis for anything that follows. 

The probability ${\cal P}$ of the detector clicking at time $\tau=0$ and position $\xi$ according to the generalised Glauber model (\ref{GlauberMixedProb}) is:
\begin{equation}
{\cal P}(\xi,0)=|\psi_\text{D}(\xi,0)|^2\langle \Psi|\hat{d}^{\dagger}\hat{d}|\Psi\rangle.\label{eqn:GlauberUDWmeet}
\end{equation} 
Of course, the Glauber model describes an idealised situation in which the measurement is performed instantaneously and at a localised position. It also ignores the fact that the apparatus itself is a complex (classical and quantum) system. Some of these considerations are taken into account in Appendix \ref{appendix1} where a model of field-apparatus interaction is considered. It is nevertheless found that up to an apparatus dependent proportionality constant (i.e., dependent on collection efficiencies etc) the single click probability is still proportional to the expectation of the operator $\hat{d}^{\dagger}\hat{d}$, i.e., the number of particles in the mode. 

Let us calculate the expectation value of the number of particles seen by an accelerating detector when the field is initialised into the Minkowski vacuum state, $|\Psi\rangle=|0\rangle_{\text{M}}$:
\begin{equation}
\label{averagenumb}
\left\langle \hat{d}^\dagger \hat{d} \right\rangle = -\left\langle (\psi^\star_\text{D},\hat{\phi}) (\psi_\text{D},\hat{\phi}) \right\rangle.
\end{equation} 
From here on the angular brackets represent taking expectation values with respect to the Minkowski vacuum state. Using (\ref{squeezedvacuum}) and (\ref{commut}) one finds: 
\begin{align}
\label{averagenum2}
\left\langle \hat{d}^\dagger \hat{d} \right\rangle 
&=\sum_{k,l}(\psi_\text{D},w_{k,I})^\star(\psi_\text{D},w_{l,I})\times\nonumber\\
&\quad\times ~_{\text{R}}\langle 0 |  \hat{S}^\dagger_{I,II}\hat{b}^\dagger_{k,I}\hat{b}_{l,I}\hat{S}_{I,II}|0\rangle_{\text{R}}\nonumber\\
&=\sum_{k}\langle \hat{n}_k\rangle|(\psi_\text{D},w_{k,I})|^2,
\end{align}
where $\langle \hat{n}_k\rangle = \sinh^2{r_k}$ is the average occupation number of a two-mode squeezed state of the plane-wave mode $k$.

Let us calculate the number of particles for the mode (\ref{modefunction}). Since we will perform several calculations in this paper using the same technique, we will go through this calculation in detail. We consider the regime, $\frac{c^2}{aL}\ll1$, so $\sigma \sim L$. If we take the cut-off at $\Lambda\sim1/L$ and make the change of variable $k'=k L$ we find:
\begin{eqnarray}
\label{averagenum3}
\left\langle \hat{d}^\dagger \hat{d} \right\rangle &=&\frac{e^{-\frac{2\pi c^2}{a L}(N-\frac{c^2 \pi}{aL})}}{4 N \sqrt{2 \pi}} \nonumber \\ \nonumber \\
& &\times\int^{\infty}_1 \frac{d k'}{k'} \frac{(N+k')^2e^{-\frac{1}{2}(k'-N-\frac{2\pi c^2}{aL})^2}}{1-e^{-\frac{2\pi k' c^2}{aL}}},
\end{eqnarray}
where we completed the square in the exponent, and took out the $k'$-independent factor. The denominator of the second factor in the integrand is approximately unity on the whole domain, furthermore in the limit $N\gg \frac{2\pi c^2}{aL}$ the integral is dominated by the contribution from $k'\sim N$. Thus, we find up to an order 1 proportionality constant:
\begin{align}
\label{averagenum4}
\left\langle \hat{d}^\dagger \hat{d} \right\rangle \sim e^{-\frac{2\pi c^2}{a L}(N-\frac{c^2 \pi}{aL})}.
\end{align}
This equation is only valid for the range of parameters satisfying the conditions above, namely $\frac{N}{2\pi}\gg \frac{c^2}{aL}\gg 1$.  As we expect, the number of particles that can be observed by the detector is exponentially suppressed by the factor $e^{-2\pi c \omega_c/a}$ where $\omega_c=\frac{c N}{L}$ is the characteristic frequency of the detected mode.

\section{Projective measurement formalism}\label{sec:Projective}

Notice that calculating the expectation value of the number of particles in the mode can be done in a purely abstract way without any reference to the detection process. In fact, working with the field theory at this abstract level of description can often be simpler and more illuminating than getting lost in the details of a specific detector model. The aim of this paper is to show that the usual phenomena of fields in non-inertial frames can be re-derived at this abstract level.

The notion of a quantum state makes sense, because in principle there could exist a device that measures it acquiring the complete information about the system encoded in the state. Appendix \ref{appendix1} shows that the abstract calculations of the previous section can also be obtained with a dynamical model of interaction between a small detector and a quantum field. In a similar way all the information extractable from a device that resolves the number of particles in a state must be describable by a set of projective operators $|n\rangle \langle n|$ associated with possible measured outcomes. This is imposed by the Hilbert space structure of quantum states.

To calculate the probability ${\cal P}(n)$ of detecting $n$ field quanta we use the projector onto the $n$-th Fock state of the mode $\hat{d}$ , that can be written in the following manner \cite{Louisell1973}: 
\begin{equation}
\label{clickprob}
{\cal P}(n) =\left\langle :\!e^{-\hat{d}^\dagger\hat{d}}\frac{(\hat{d}^{\dagger}\hat{d})^n}{n!}\!:  \right\rangle.
\end{equation}
In the weakly coupled regime the above expression can be derived from a dynamical model described in the Appendix \ref{appendix1}, with an extra modification of $\hat{d}^\dagger \hat{d}$ being replaced with $\eta\hat{d}^\dagger \hat{d}$, where $\eta\leq 1$ \cite{Mandel}. The new parameter $\eta$ describes the efficiency of the detection and is related to the coupling strength between the systems.

The probability distribution \eqref{clickprob} can be related to the characteristic function $Z(\lambda)$ via the Fourier transform:
\begin{eqnarray}
\label{characteristic}
{\cal P}(n) &=& \int_0^{2\pi}\frac{\text{d}\lambda}{2\pi} e^{-i\lambda n}Z(\lambda).
\end{eqnarray}
Here
\begin{eqnarray}
Z(\lambda) &=& \sum_{n=0}^\infty e^{i\lambda n}{\cal P}(n) = \left\langle :\!e^{(e^{i\lambda}-1)\hat{d}^\dagger\hat{d}}\!:  \right\rangle\nonumber \\ 
&=&\sum_n \frac{(e^{i\lambda}-1)^n}{n!}\,_{\text{R}}\langle 0 |\hat{S}^\dagger_{I,II} \hat{d}^{\dagger n}\hat{d}^{n}\hat{S}_{I,II}|0\rangle_{\text{R}} 
 \nonumber \\
&=& \sum_n \frac{(e^{i\lambda}-1)^n}{n!}\,_{\text{R}}\langle 0 | \left[\sum_k\sinh r_k (\psi_\text{D},w_{k,I})^\star\hat{b}_{k,II}\right]^{n}\nonumber\\
&~&\times\left[\sum_l\sinh r_l(\psi_\text{D},w_{l,I})\hat{b}^\dagger_{l,II}\right]^{n} |0\rangle_{\text{R}} \nonumber \\ \nonumber \\
&=&  \sum_n (e^{i\lambda}-1)^n\langle \hat{d}^\dagger \hat{d}\rangle^n = \frac{1}{1-(e^{i\lambda}-1)\langle \hat{d}^\dagger \hat{d} \rangle},
\end{eqnarray} 
where the multinomial expansion was used to obtain the last line and the mean particle number $\langle \hat{d}^\dagger \hat{d} \rangle$ is given by Eq.~\eqref{averagenum2}. Through \eqref{characteristic} we find the excitation statistics:
\begin{equation}
\label{clickprob2}
{\cal P}(n) = \frac{\langle \hat{d}^\dagger \hat{d} \rangle^n}{(1+\langle \hat{d}^\dagger \hat{d} \rangle)^{1+n}}
\end{equation}
for the Minkowski vacuum state $|0\rangle_{\text{M}}$.

\section{Determining the Unruh temperature}\label{sec:Unruh}

One might wonder what temperature is associated with the above statistics. For a thermalised state of a harmonic oscillator the temperature is defined by the relation $\langle \hat{d}^\dagger \hat{d}\rangle =\left(e^{E/kT} -1\right)^{-1}$, where $E$ is the energy of a single excitation of the field mode. However in general the mode $\psi_{\text{D}}$ has a frequency spread so it does not have well defined energy, therefore the temperature is also not well defined. Nevertheless, one can define a temperature estimator $T_\text{est}$ by replacing $E$ with the expectation value of the energy of a single excitation of the field mode,  $E \rightarrow \,_\text{R}\langle 0|\hat{d} \hat{H}_{\text{R}}\hat{d}^{\dagger}|0\rangle_{\text{R}}=  \sum_{k}\hslash\omega_k | (\psi_\text{D},w_{k,I})|^2$, where $\hat{H}_{\text{R}}=\sum_k \hbar \omega_k \hat{b}_{k,I}^{\dagger}\hat{b}_{k,I}+(I\leftrightarrow II)$ is the generator of proper time translations in the Rindler frame and $\omega_k$ are Rindler frequencies. For a detector moving with proper acceleration, $a$, this temperature estimator is equal to:\begin{equation}
\label{temperature}
kT_\text{est} = \frac{\sum_{k}\hslash\omega_k |(\psi_\text{D},w_{k,I})|^2}{\log\left(1 + \langle \hat{d}^\dagger\hat{d} \rangle^{-1}\right)}.
\end{equation}
Here we find that the temperature observed depends on the proper acceleration of the detector, $a$, and the shape of the mode $\psi_{\text{D}}$.

For the choice of mode (\ref{modefunction}), we find that the temperature estimator \eqref{temperature} reduces approximately (when $\frac{N}{2\pi}\gg \frac{c^2}{aL}\gg 1$) to:
\begin{equation}
kT_\text{est}\approx\frac{\hslash a}{2\pi c}\frac{1}{1-\pi c^2/aL N}.
\end{equation} 
Note that the temperature estimator deviates from the Unruh temperature formula \cite{Unruh1976} for finite $N$. This stems from the fact that the energy of the mode is not well-defined. However, for large $N$, which corresponds to a peaked energy spectrum, we recover the celebrated Unruh result \cite{Unruh1976}.

\section{Non-locality of the vacuum state} \label{sec:vacent}

In order to extract entanglement from the vacuum state we need more than just one detector. Consider two identical detectors, one moving with proper acceleration $a>0$ in region $I$ and coupled to a mode $\psi_{\text{D},I}$, and the other moving with acceleration $-a$ in region $II$ and coupled to $\psi_{\text{D},II}$. The corresponding annihilation operators are $\hat{d}_I$ and  $\hat{d}_{II}$ respectively and both detectors are causally disconnected. One can generalise the calculation \eqref{averagenum2} to determine the average product of the particle content detected in the two modes and obtain the result: 
\begin{equation}
\label{classicalcorrelations}
\left\langle \hat{d}^\dagger_I \hat{d}_I \hat{d}^\dagger_{II} \hat{d}_{II} \right\rangle = \left\langle \hat{d}^\dagger_I \hat{d}_I \right\rangle \left\langle \hat{d}^\dagger_{II} \hat{d}_{II} \right\rangle +\left| \left\langle \hat{d}_I \hat{d}_{II} \right\rangle \right|^2,
\end{equation}
where the first two terms on the RHS of the equation are given by \eqref{averagenum2} and the last term is equal to:
\begin{eqnarray}
\label{correlationterm}
\left\langle \hat{d}_I \hat{d}_{II} \right\rangle &=& \sum_k \sqrt{\langle \hat{n}_k \rangle(1+\langle \hat{n}_k \rangle)}\nonumber \\ \nonumber \\
& &\times(\psi_{\text{D},I},w_{k,I})(\psi_{\text{D},II},w_{k,II}).
\end{eqnarray}
We find that the measurement outcomes carried out by the detectors are correlated and the correlations are determined by the magnitude of expression \eqref{correlationterm}. In order to prove the non-locality of these correlations we note that a two party symmetric Gaussian state is entangled if \cite{Duan2000}: 
\begin{equation}
\label{DuanCrit}
\left\langle\left(\hat{x}_I-\hat{x}_{II}\right)^2 \right\rangle\left\langle\left(\hat{p}_I+\hat{p}_{II}\right)^2 \right\rangle < 1,
\end{equation}
where $\sqrt{2}\hat{x}_\sigma = \hat{d}_\sigma + \hat{d}_\sigma^\dagger$ and $\sqrt{2}i\hat{p}_\sigma = \hat{d}_\sigma - \hat{d}_\sigma^\dagger$ for $\sigma\in\{I,II\}$.  

In order to detect entanglement we need to re-program our detectors to perform projective measurements corresponding to the hermitian quadrature operators $\hat{x}_\sigma$ or $\hat{p}_\sigma$ instead of projecting in the Fock basis. Typically, such a measurement is realised by means of homodyne detection with the use of an auxiliary beam of light \cite{Dragan2001}. In the present work we do not study the details of such setup \cite{Downes2012} assuming that our device carries out an ideal measurement of the quadratures. We find that the left-hand side of the inequality \eqref{DuanCrit} is equal to
$\left(  1+ \left\langle \hat{d}^\dagger_I \hat{d}_I \right\rangle + \left\langle \hat{d}^\dagger_{II} \hat{d}_{II} \right\rangle - 2\text{Re} \left\langle \hat{d}_I \hat{d}_{II} \right\rangle \right)^2$ 
and again it turns out that the presence of the entanglement is dictated by the same term \eqref{correlationterm} that was responsible for the existence of correlations between the detectors' counts.
\begin{figure}
\begin{center}
\includegraphics[width=\columnwidth]{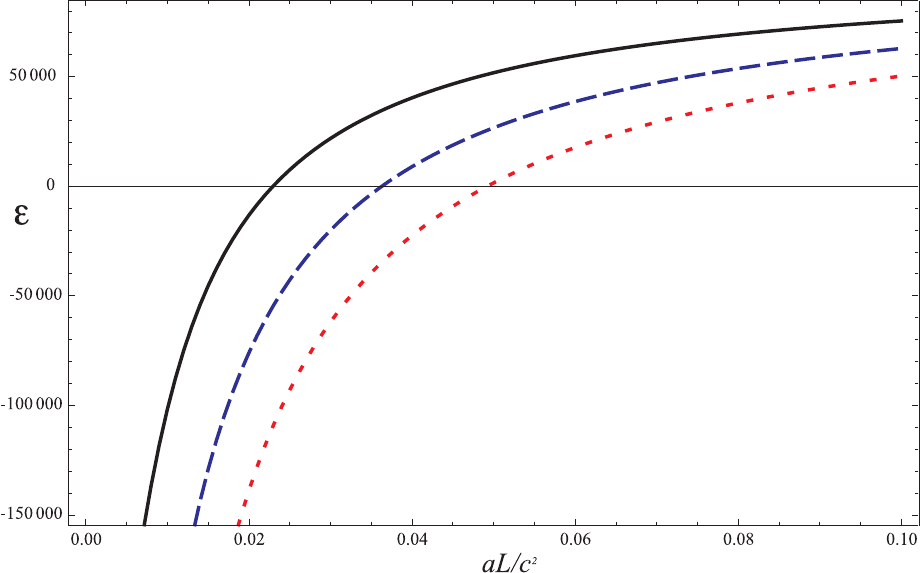}
\end{center}
\caption{\label{entanglement}Entanglement estimator ${\cal E}$ as a function of the dimensionless parameter $\frac{aL}{c^2}$ for $N=800$ (solid line), $N=1200$ (dashed line) and $N=1600$ (dotted line).}
\end{figure}
For a pair of identical, counter-accelerating detectors we have $\left\langle \hat{d}^\dagger_I \hat{d}_I \right\rangle = \left\langle \hat{d}^\dagger_{II} \hat{d}_{II} \right\rangle$. By this criterion we have checked that for any finite acceleration the state measured by the detectors is entangled. Knowing this, we can now quantify this entanglement extracted by the detectors by introducing the entanglement estimator ${\cal E}$:
\begin{equation}
\label{estimator}
{\cal E} = \log\left| \left\langle \hat{d}^\dagger_{I} \hat{d}_{I} \right\rangle - \text{Re} \left\langle \hat{d}_I \hat{d}_{II} \right\rangle \right| + C
\end{equation}
where $C$ is an arbitrary real constant factor. Here ${\cal E}\rightarrow-\infty$ when the detectors state is separable and grows monotonically as the degree of quantum entanglement is increased. For $\psi_{\text{D},I}$ given by \eqref{modefunction} and $\psi_{\text{D},II}$ given by the transformation $\xi' = -\xi$ of \eqref{modefunction} we have $(\psi_{\text{D},II},w_{k,II})=(\psi_{\text{D},I},w_{k,I})$. Taking the limit of $\frac{N}{2\pi}\gg \frac{c^2}{aL}\gg 1$ the estimator can be approximated by:
\begin{equation}
\label{estimatorapprox}
{\cal E} \approx  -\frac{\pi c^2}{aL}\left(N-\frac{\pi c^2}{2aL} \right)+ C.
\end{equation}
We know that the extracted entanglement must vanish (${\cal E}\rightarrow\nolinebreak-\infty$) as $a\rightarrow 0$, although for fixed $N$ our approximations are not valid in this limit. In Fig.~\ref{entanglement} we plot the dependence of the entanglement ${\cal E}$ \eqref{estimator} on the dimensionless quantity $\frac{aL}{c^2}$ for several values of $N$, finding good agreement with the approximated result \eqref{estimatorapprox}. We find that the largest amount of entanglement can be extracted from the Minkowski vacuum state when the proper lengths and proper accelerations of the detectors are large and the frequency numbers $N$ are low.

\section{Discussion and conclusions} \label{sec:conclusion}
Our results may help to understand the nature of the correlations between localised measurements of the quantum vacuum studied in the context of algebraic QFT \cite{Summers1985}. It is also interesting to compare our findings with the results of \cite{Zych2010}, where entanglement between two localised, orthogonal modes was observed in the inertial frame. This entanglement was shown to decay with the spatial separation of the modes. On the other hand, we find that no entanglement can be extracted from inertial detectors. This difference can be attributed to the fact that our detectors are probing modes that consist only of positive frequencies and the authors of \cite{Zych2010} consider modes that have no such restriction.

The main idea of this paper is to emphasise that for single mode measurements the outcomes of photo-count experiments in the accelerating frame can be described by projective operators operating only on the single mode subspace. In this way calculations can be performed at an abstract level without making any reference to the actual detection processes. The versatility and power of this approach is demonstrated by the analytic expressions we were able to re-derive for the standard phenomena in non-inertial frames i.e., the Unruh effect and extraction of vacuum entanglement. Although here we restricted our attention to measurements of the occupation number of the mode (see Section \ref{sec:particles}) and quadrature measurements (see Section \ref{sec:vacent}), in principle any observable in question can be chosen at will by the experimentalist.


Although it is both mathematically and conceptually advantageous to work at the abstract level of projective operators, one must keep in mind the underlying physical processes. For example, the state of the measured system is completely destroyed during the measurement process (photons are absorbed) and the physical measurement is not instantaneous and not taking them into account can lead to problems \cite{Sorkin1993}. In Appendix \ref{appendix1} we showed by taking the interaction time into account that under reasonable conditions the results of the generalised Glauber model can be validated.
 
The model presented here can be readily applied to other types of motion, as well as curved spacetimes in regions where time-like Killing vectors exist. Further research is currently being undertaken to use this approach to investigate localised entangled states, instead of the Minkowski vacuum. The principles laid out in this work can be generalised to other types of fields, and it is expected that new insights into the nature of entanglement in accelerated frames are to be found in this approach.

\acknowledgements 
We thank P. Alsing, J. Louko, M. Montero, T. Ralph, and M. Del Rey for very useful discussions. A.~D. thanks for the financial support to EPSRC [CAF Grant EP/G00496X/2] and the Polish Ministry of Education.
\appendix\section{}
\label{appendix1}
In this appendix we consider a dynamical interaction model for the photon absorption processes occurring at $D$ in Fig.~\ref{scheme} and show that the model reproduces the general behaviour of a generalised Glauber model (\ref{eqn:GlauberUDWmeet}).  

In idealised photo-detectors electrons in bound potential wells are excited into free states by the absorption of photons, see for example \cite{Mandel}. In this approach the interaction of the photon with the electron is calculated using a non-relativistic Schr\"{o}dinger equation for the electron and using the minimal coupling substitution $p\rightarrow p-e A(x,t)$ resulting in an interaction term, 
\begin{equation}\label{pdotAinteraction}
H_I=\frac{e}{m} \hat{p} \cdot \hat{A}(r_0,t).
\end{equation}
 Since the transformation of this setup into the accelerated frame is complicated by many factors (including relativistic considerations, and the effects of acceleration on the binding material properties etc) it is more practical to consider a detector with subsystems interacting with the field via the interaction Lagrangian: 
\bea\label{UDW}
L[x(\tau)]= \mu \hat{m}(\tau) \hat{\phi}[x(\tau)]
\eea
where $x(\tau)$ is the classical path followed by the detector, $\tau$ is the proper time along the path and $\hat{m}$ is a monopole moment having discrete energy levels. The time evolution of the monopole moment can be written, $\hat{m}(\tau)=e^{i\hat{H}_0 \tau/\hbar}\hat{m}(0) e^{-i\hat{H}_0 \tau/\hbar}$ where $\hat{H}_0$ is the free Hamiltonian for the monopole. This is of course just an Unruh-DeWitt interaction \cite{Birrell1982}. For an inertial path equations (\ref{pdotAinteraction}) and (\ref{UDW}) are remarkably similar. We can even approximate the continuum of energy states in the photo-electric case (above the binding energy $ > \hbar \omega_0$) by discrete energies of the monopole. For example, we could assume that the energies of the monopole have a large gap from ground to first excited state representing the binding energy of the electron potential then smaller perhaps equally spaced gaps thereafter. One would then be able to reproduce the standard theory of photo-detection using this model \cite{Mandel}. Furthermore, since the interaction transforms simply as a scalar under coordinate transformations, one would also expect to be able to perform the equivalent calculations in the accelerated frame. It is this direction we will now pursue.

We assume the transition to a higher energy state of the monopole, as in the photo-electric model, results in a current or cascade process that magnifies the occurrence of the quantum transition into a classical measurement. However, we don't model the quantum to classical process here.

We assume the "binding energy" for this model is of the typical energy scale of the mode of the field we measure. For the Gaussian in (\ref{modefunction}) we would have $E_1-E_0\sim \frac{N \hbar c}{L}$. 
 
The initial field state is taken to be the mixed state (\ref{mixedstate}) and the monopole is initialised into the ground state $\rho_m(\tau_0)=|E_0\rangle\langle E_0|$, so that the initial state of the system (monopole plus single field mode) is given by the product state $\hat{\rho}(\tau_0)=\hat{\rho}_m(\tau_0)\otimes \text{Tr}_{\perp \psi_\text{D}} |\Psi\rangle \langle \Psi |$. From the outset we work in the accelerated frame with the detector positioned at $\xi=\xi_0$. In this frame the system will be assumed to undergo evolution according to the operator of infinitesimal displacements of proper time. The different inertial and accelerating photo-count statistics will then be seen to arise because of the different proper time parameters under which the Hamiltonian systems evolve respectively.  The probability of finding the system in some orthogonal state $\hat{\rho}_f=|E,\chi\rangle\langle E,\chi|$ after an interaction time $T$ can be calculated using a perturbation expansion of the equations of motion for the density matrix. The second order approximation is given by \cite{Mandel}:
\begin{align}\label{secordpert}
{\cal{P}}(&\xi_0, \tau_0; T)=\frac{\mu^2 }{\hbar^2}\int^{\tau_0+T}_{\tau_0}d\eta_1\int^{\eta_1}_{\tau_0} d\eta_2\times \\\nonumber
 &\langle E,\chi| m(\eta_1) \hat{\phi}(\xi_0,\eta_1)\hat{\rho}(\tau_0)m(\eta_2) \hat{\phi}(\xi_0,\eta_2)|E,\chi\rangle +c.c,
\end{align}
To this order the only processes that occur are those in which a single photon is either created or annihilated. In analogy with electrons excited in photo-diodes we assume a classical click of the detector occurs when the monopole makes a transition into any of the excited states. Therefore, to calculate the probability of a click we sum over all possible final energy states of the monopole. Furthermore, since when we obtain a click we don't know which state the field has transitioned into the total single click probability is found by summing over all possible final states of the field. We then arrive at the probability for a single detector click in a time $T$ given by: 
 \begin{widetext}
 \bea
{\cal{P}}(\xi_0, \tau_0; T)=\frac{\mu^2}{\hbar^2} \sum_E |\langle E|m(0)|E_0\rangle|^2 \mathcal{F}(E-E_0)
\eea
where 
\bea\label{responsefn}
\mathcal{F}(E-E_0)=\int^{\tau_0+T}_{\tau_0}d\eta_1\int^{\eta_1}_{\tau_0} d\eta_2 e^{i (E-E_0) (\eta_1-\eta_2)/\hbar} \text{Tr}[ \hat{\phi} (\xi_0,\eta_2)\hat{\phi} (\xi_0,\eta_1) Tr_{\perp \psi_\text{D}} |\Psi\rangle \langle \Psi |] +c.c.
\eea 
\end{widetext}
Consider the terms in equation (\ref{rindlerdecomp}) 
since the accelerated worldline $x_0(\tau)=(\xi_0, \tau)$ lies completely in region I, the region II operators do not contribute to the monopole interactions (\ref{UDW}). Therefore,
\begin{align}
\hat{\phi}[x(\tau)]=
\int dk\hat{b}_{k,I}w_{k,I}(\xi_0,\tau)+\hat{b}_{k,I}^{\dagger}w_{k,I}(\xi_0,\tau)^\star\equiv\hat{\psi}_I(\xi_0,\tau),
\end{align}
and using the expansion:
\begin{widetext}
\begin{align}
\hat{\psi}_{I}(\xi_0,\eta_2)\hat{\psi}_{I}(\xi_0,\eta_1)
=\quad:&\hat{\psi}_{I}(\xi_0,\eta_2)\hat{\psi}_{I}(\xi_0,\eta_1):+\int\frac{dk}{4 \pi |k|} e^{-i |k|c (\eta_2-\eta_1)},\label{novacsep}
\end{align}
equation (\ref{responsefn}) can be rewritten:
\begin{align}\label{responsefnseparated}
\mathcal{F}(E-E_0)=\int^{\tau_0+T}_{\tau_0}&d\eta_1\int^{\eta_1}_{\tau_0} d\eta_2 e^{i (E-E_0) (\eta_1-\eta_2)/\hbar} \times\nonumber\\&\left
(\text{Tr}\left\{:\hat{\psi}_{I}(\xi_0,\eta_2)\hat{\psi}_{I}(\xi_0,\eta_1):\text{Tr}_{\perp \psi_\text{D}} |\Psi\rangle \langle \Psi |\right\}+\int\frac{dk}{4 \pi |k|} e^{-i |k|c (\eta_2-\eta_1)}\right) +c.c.
\end{align}
\end{widetext}
It is useful to separate $\mathcal{F}(E-E_0)=\mathcal{F}_{n.o}(E-E_0)+\mathcal{F}_{vac}(E-E_0)$ into a normally ordered part and a vacuum fluctuation contribution (that is independent of the state). First we focus on the vacuum fluctuation contribution. Defining $\eta=\eta_1-\eta_2$, we find:
\begin{align}\label{vacfluccontrib}
\mathcal{F}_{vac}(E-E_0)=\int \frac{dk}{4\pi |k|} \int^{T}_{-T} d\eta (T -|\eta|) e^{i((E-E_0)/\hbar+|k|c)\eta}.
\end{align}
Consider the innermost integrand. Since the time of measurement is much larger than the typical timescale of a transition, $T\gg\frac{\hbar}{E_1-E_0}$, for all wave numbers $k$ the phase oscillates rapidly over the $\eta$ integration domain. The inner integral is then approximately equal to $2\pi \delta((E-E_0)/\hbar+|k|c)$. Since this is zero for all $k$ the vacuum contribution is negligible. 

Now to evaluate the normally ordered field contribution we use the expansion:
\begin{equation}
\label{ddecomp}
\hat{\phi}(\xi,\tau) = \hat{d}\psi_\text{D}(\xi,\tau)+ \hat{d}^\dagger\psi_\text{D}^\star(\xi,\tau)+\hat{\phi'}.
\end{equation}
In normal order all the annihilation operators are to the right of the creation operators, however, only the mode $\psi_D$ is occupied in the state accessible to the detector, $\text{Tr}_{\perp \psi_\text{D}} |\Psi\rangle \langle \Psi |$. Therefore, no contributions to (\ref{responsefn}) can arise from the $\hat{\phi}'$ part of the field operator since those terms will be zeroed by acting on the orthogonal subspace vacuum state, therefore:
\begin{align}
&\text{Tr}\left\{:\hat{\psi}_{I}(\xi_0,\eta_2)\hat{\psi}_{I}(\xi_0,\eta_1):\text{Tr}_{\perp \psi_\text{D}} |\Psi\rangle \langle \Psi |\right\} \nonumber\\
&=\psi_\text{D} (\xi_0,\eta_2)\psi_\text{D} (\xi_0,\eta_1)\langle\Psi| \hat{d}^2|\Psi\rangle\nonumber\\
&\quad+\psi_\text{D}^\star (\xi_0,\eta_2)\psi_\text{D} (\xi_0,\eta_1)\langle\Psi| \hat{d}^{\dagger}\hat{d}|\Psi\rangle +c.c.\label{nodecomp}
\end{align}
To analyse $\mathcal{F}_{n.o}(E-E_0)$ further, it is useful to decompose $\psi_\text{D}$ into Rindler frequencies:
\bea
\psi_\text{D}(\xi,\tau)&=&\int(\psi_\text{D},w_{k,I}) w_{k,I}(\xi,\tau)dk.
\eea
Consider the contribution to (\ref{responsefnseparated}) from the first term in (\ref{nodecomp}), let $\Delta\omega=\frac{E-E_0}{\hbar}$ then:
\begin{align}
\int^{\tau_0+T}_{\tau_0}&d\eta_1\int^{\eta_1}_{\tau_0} d\eta_2 e^{i (E-E_0) (\eta_1-\eta_2)/\hbar}\psi_\text{D} (\xi_0,\eta_2)\psi_\text{D} (\xi_0,\eta_1)\nonumber\\
=&\int\frac{dk dk' (\psi_\text{D},w_{k,I})(\psi_\text{D},w_{k',I})}{4 \pi \sqrt{|k| |k'|}}e^{i(\xi_0(k+k')-\tau_0(|k|+|k'|)}\nonumber\\
&\times \int^{T}_{0}d\eta_1e^{i (\Delta\omega-|k'|)\eta_1}\int^{\eta_1}_{0} d\eta_2 e^{-i (\Delta\omega+|k|)\eta_2}.\label{counterrotterm}
\end{align}
Since the $\psi_D$ mode has a peaked frequency approximately equal to the typical transitional frequency, $\omega_c\sim \Delta\omega$, the innermost integral oscillates at double the optical frequency. But the measurement time is assumed to be many optical periods long (i.e., $T\sim L/c=\frac{N}{\omega_c}$), therefore this term only makes a small contribution to the total integral. By analysing all the other terms in a similar way, it can be seen that the dominant contribution comes from the second term in (\ref{nodecomp}). One may recognise this step as the Rotating Wave Approximation, however this is a different approximation to the usual RWA in the inertial frame because the frequencies here are Rindler frequencies. One then obtains:
\begin{widetext}
\bea\label{approxresponsefn}
\mathcal{F}(E-E_0)&=&\langle\Psi| \hat{d}^{\dagger}\hat{d}|\Psi \rangle \int^{T}_{0}d\eta_1\int^{\eta_1}_{0} d\eta_2 e^{i (E-E_0) (\eta_1-\eta_2)/\hbar} \psi_\text{D}^\star (\xi_0,\eta_2+\tau_0)\psi_\text{D} (\xi_0,\eta_1+\tau_0)+c.c.
\eea 
Note that all the dependence on the state of the field, $|\Psi \rangle$, has completely factored out of the integral. The probability of a single click at the spacetime point $(\xi_0,\tau_0)$ (over a measurement duration $T$) then takes the form:
\bea\label{pclick}
{\cal{P}}(\xi_0,\tau_0;T)= \alpha(\xi_0,\tau_0;T)\langle\Psi| \hat{d}^{\dagger}\hat{d}|\Psi \rangle,
\eea
where 
\begin{align}
\alpha(\xi_0,\tau_0,T)&=\frac{\mu^2}{\hbar^2} \sum_E |\langle E|m(0)|E_0\rangle|^2 \int^{T}_{0}d\eta_1\int^{\eta_1}_{0} d\eta_2 e^{i (E-E_0) (\eta_1-\eta_2)/\hbar} \psi_\text{D}^\star (\xi_0,\eta_2+\tau_0)\psi_\text{D} (\xi_0,\eta_1+\tau_0)+c.c.
\end{align}
\end{widetext}
We now consider two measurement scenarios. Consider first the case when the measurement time $T$ is much shorter than the period of the entire pulse but still longer than the binding period, $\frac{\hbar}{E_1-E_0}\ll T\ll L/c$. In this regime, the measurement time is too short to resolve the spectral line-width of the wave-packet and so the quasi-monochromatic approximation can be used:
\bea
\psi_\text{D} (\xi_0,\eta_1+\tau_0)\sim\psi_\text{D} (\xi_0,\tau_0)e^{-i \omega_c\eta_1}
\eea
where $\omega_c=\frac{c N}{L}$. In this case, $\alpha(\xi_0,\tau_0,T)\propto T$. Therefore, the single particle absorption probability scales linearly with the measurement duration.  

The second case we consider is when the measurement time is taken to be greater than the time it takes the single mode to traverse the photo-detector, $T>L/c$. Then frequencies other than the peak frequency will contribute to the detection probability. In this case equation (\ref{pclick}) must be interpreted as the probability of a single count per shot (unit time $T$). 
One can build up photo-count statistics by preparing an ensemble of identical systems and performing a single shot measurement (of time $T$) on each system in the ensemble. 
%

We conclude this appendix by noting that taking $|\Psi\rangle$ to be the Minkowski vacuum state then agrees up to a device-dependent proportionality constant with the same expression arrived at using the generalised Glauber mode of detection (\ref{eqn:GlauberUDWmeet}).

\end{document}